# On the question of the electromagnetic momentum of a charged body

V. B. Morozov

*The incorporation of a relativistic momentum of a nonelectromagnetic nature into macroscopic problems of electrodynamics obviates the lack of correspondence between the electromagnetic mass and the electromagnetic momentum of macroscopic bodies, allowing, in particular, the resolution of the well-known `4/3 paradox'.*

1. Almost immediately after the discovery of cathode rays, the idea was proposed that the electromagnetic field of the electron has a mass and indeed that the electron mass has an entirely field origin. But this led to the problem noted by J J Thomson already in 1881 [1] – `the troublesome and puzzling factor 4/3' [2] occurring in the electromagnetic inertia in the charged sphere model of the electron. The consideration of a relativistically contracted sphere by Lorentz led to the same result. At various times, Henri Poincare, Albert Einstein, and M Abraham, to name but a few, addressed the question of the electromagnetic mass and momentum. Today, the classical theory of the electron has a merely historical interest. A problem well outside mainstream theory, the electromagnetic mass in the classical model of the electron was regarded rather as an annoying misunderstanding or as an excuse for discussing the incompleteness of classical electrodynamics. Still, even today, the electromagnetic mass of a charged body treated as a problem in classical electrodynamics may be interesting.

With the example of simple problems given below, we attempt to demonstrate the consistency of the relativistic theory of the motion of charged bodies.

2. We recall the well-known paradoxical solution to the problem of the electromagnetic mass of a uniformly charged sphere [2-5]. The starting point here is the expression for the momentum density of the field,

$$\mathbf{g}_f = \varepsilon_0 \mathbf{E} \times \mathbf{B}.$$

which, when integrated over the volume, yields the following expression for the field momentum of a charged sphere of radius $a$ moving with a uniform velocity $v$ for small $v/c$:

$$\mathbf{P}_f = \frac{4}{3} \frac{q^2}{8\pi\varepsilon_0 a} \frac{\mathbf{v}}{c^2}, \tag{1}$$

where $q$ is the charge. On the other hand, from the energy density of the electric field,

$$\rho c^2 = \frac{\varepsilon_0}{2} E^2,$$

the total electromagnetic mass of the field of a charged sphere is found as

$$m_{em} = \frac{q^2}{8\pi\varepsilon_0 a} \frac{1}{c^2}. \tag{2}$$

Comparing Eqn (1) and Eqn (2) shows that, contrary to the expectation, the masses of the electromagnetic field are not identical. The mass ratio 4/3, which gave the name to the paradox, has until recently been generally explained by attributing a no electromagnetic mass to the `electron' (a charged sphere). We note here that a paradoxical solution arises not only when comparing the electromagnetic masses of the `electron' but also for all similar problems. For example, for a uniform change within a sphere, this ratio is 8/5 [2].

Any classical model of the electron is inherently unstable and requires that some forces be introduced to compensate the Coulomb repulsion. Poincare suggested the presence of scalar *Poincare tensions* in the electron, capable of compensating Maxwell's tensions: they were supposedly due to a certain massive field and were designed to account for the missing mass of the electron. This formal approach led to nothing, however. Ultimately, the paradoxical nature of this solution gave rise to speculations on the inconsistency of classical electrodynamics and/or the special theory of relativity. Today, with a century of history behind it, the problem is regarded as an unresolvable paradox [2, 5]. Amazingly, no notice has ever been taken to the proof of momentum-energy balance for charged bodies, a result published in Becker's textbook as far back as 1933 [6][1]. Becker showed that the factor 4/3 in the momentum of a charged sphere is due to an additional momentum, which, although of a no electromagnetic origin, is related to Maxwell's tensions. In another finding, he showed that the problem could be solved consistently by assuming that the force densities due to Maxwell's tensions and those due to elastic tensions compensate each other. This, in fact, is tantamount to the energy and moment conservation theorem (for closed systems) being generalized both in terms of its formulation and its proof. This is probably why Becker's result remained unnoticed for three quarters of a century.[2]

That the energy-momentum conservation is satisfied is taken for granted for a system of coupled charges. As we see below, an unexpected fact is that the momentum of the system is redistributed, and unavoidably so, between the field momentum and the mechanical momentum, whereas the mass of the field and the mechanical mass show nothing of the kind.

3. We first consider the simplest one-dimensional problem of a charged planar capacitor moving in the *x* direction perpendicular to its plates A and B. The space between the plates of the moving capacitor contains energy, to which we must assign a mass. However, this mass is not contained in the electromagnetic momentum of the capacitor: we note that the electric field **E** is parallel to the velocity of the capacitor $\mathbf{v} = (v_x, 0, 0)$, and that the Poynting vector and the electromagnetic momentum are both zero. This even more blatant inconsistency between the electromagnetic momentum and the electromagnetic mass could well be called the zero or 0/1 paradox.

---

[1] Reference [6] was brought to our attention by an anonymous referee of the first version of this paper.

[2] To our knowledge, the only mention of this method in relation to the problem of electromagnetic mass is by the editors of the Russian translation of Pauli's book [3].

The special theory of relativity, when applied consistently, helps solve this problem. With the energy density of the capacitor electric field denoted by $w$, the pressure on the surface of the dielectric is

$$p = \frac{dw}{dx}. \tag{3}$$

Because the planes A and B move at the velocity $v_x$, the pressure p performs work on both planes of the dielectric, such that the power per unit area is $v_x p$ for plane A and $-v_x p$ for B. At first sight, this does not change anything because, when summed over the entire system, both the force and work are zero. But we observe an energy transfer from plane A to plane B. Hence, the energy flow density is $S = v_x p$. This quantity is related to the momentum density (also known as the hidden momentum) by

$$g_x = \frac{1}{c^2} S, \tag{4}$$

which is a direct consequence of Einstein's mass-energy equivalence [7]. We let $L$ be the length of the dielectric layer in the laboratory frame. Then integrating Eqn (4) and using Eqn (3), we find the capacitor momentum per unit area as

$$G_x = \frac{v_x}{c^2} \int_0^\delta \frac{dw}{dx} dx = v_x \frac{wL}{c^2}. \tag{5}$$

This result is in exact correspondence with the mass of the capacitor (also per unit area)

$$m_f = \frac{1}{c^2} \int_0^\delta w\, dx = \frac{wL}{c^2}. \tag{6}$$

We note here that the factor $\gamma = (1 - v^2/c^2)^{-\frac{1}{2}}$ is already incorporated in the capacitor thickness $L$. Quantities (5) and (6) are found from the electromagnetic energy density. There was no need for us to calculate the energy of the electromagnetic field. The essential point, however, is that the momentum is entirely that of a mechanical system, i.e., of a continuum medium, even though the parameters of this medium do not enter the final result. In summary, as this example shows, it is only by simultaneously considering the electromagnetic field and the mechanical system that the motion of a system of charged bodies can be correctly described.

4. In considering a closed system of charged bodies, it is necessary that the energy-momentum tensor of the electromagnetic field be augmented by the energy-momentum tensor of the continuous body [8]

$$T^{ik} = \begin{pmatrix} \rho c^2 & cg_x & cg_y & cg_z \\ cg_x & t_{xx} & t_{xy} & t_{xz} \\ cg_y & t_{yx} & t_{yy} & t_{xz} \\ cg_z & t_{zx} & t_{zy} & t_{zz} \end{pmatrix},$$

where $\rho$ is the density, $cg_i$ are the energy flow components, $g_i$ are the momentum components, and $t_{\alpha\beta}$ is the mechanical stress tensor. In particular, for a fixed homogeneous sphere or simply for a drop of liquid with the internal pressure $p$, the energy-momentum tensor has the form [9]

$$\begin{pmatrix} \rho c^2 & 0 & 0 & 0 \\ 0 & p & 0 & 0 \\ 0 & 0 & p & 0 \\ 0 & 0 & 0 & p \end{pmatrix}. \tag{7}$$

A Lorentz transformation of the tensor yields a relation for the momentum density expressed in terms of the reference frame velocity $v_x$, and the density and tensions in the comoving system [8]:

$$g_x = \gamma^2 \rho^0 v_x + \frac{\gamma^2}{c^2} t^0_{xx} v_x, \qquad g_y = \frac{\gamma}{c^2} t^0_{xy} v_x, \qquad g_z = \frac{\gamma}{c^2} t^0_{xz} v_x, \tag{8}$$

where $t_{xx} v_x$, $t_{xy} v_x$, $t_{xz} v_x$ are the components of the energy flow density. A charge on the surface of charged bodies produces mechanical stresses within them, implying the existence of a nonzero nonelectromagnetic momentum density, which does not vanish when integrated over the volume of the system. The momentum density is ultimately related to charges in the system, not to the density of the medium it contains. Namely, the symmetric nature of the charged sphere problem implies that a nonzero contribution to the total momentum comes only from $g_x$.

5. If we consider an elementary rectangular parallelepiped with sides $dxdydz$ inside a medium, whose faces are acted upon along the *x* axis by two equal-magnitude, oppositely directed forces $t_{xx}dydz$, then, similar to the capacitor problem above, the motion of this volume element along the *x* axis results in the applied forces performing the work $t_{xx}v_x dydz$ and $-t_{xx}v_x dydz$ per unit time. We note that the energy flow transfers energy at the rate $t_{xx}v_x dxdydz$ per unit time over a distance $dx$. We also note that if the forces acting on the opposite faces are opposite, then the energy flow and velocity vector have the same direction; if the forces are outward, then the energy flow is in the opposite direction to the vector **v**. Accordingly, as with Eqn (4), the sign of the relativistic momentum of an element of the medium is dependent on that of the pressure.

6. We note the nonzero nature of the quantity **E · j** (with **j** being the current density) that enters the energy balance equation (Poynting theorem) in the problems we are considering. For a charged sphere, this means that the field does work (on the sphere, in our case). For positive **E · j**, energy is transferred from the field to the sphere, while for negative Ej, it is transferred in the opposite direction. Because the picture is symmetric, the contribution of **E · j** to the energy balance is zero and might, seemingly, be neglected. But inside the sphere, there is an energy flow that is transferred without the mediation of the field (see Fig. 1). This energy flow – and this is suggested by the relativistic energy-mass equivalence principle – is equivalent to momentum. Or, equivalently, the stress tensor components of the sphere have the momentum density defined by Eqn (8).

7. We consider a sphere centered at the origin of a Cartesian coordinate system. According to Eqn (8), a volume element of the sphere – a disk of thickness dx formed by two planes parallel to the *yz* plane and centered at *x* – has the momentum

$$dp_x = dx \frac{v_x}{c^2} \int_S t_{xx} dydz = dx \frac{v_x}{c^2} T,$$

where *s* is a surface element, $\pi(a^2 - x^2)$ is the area, and *T* is the force acting on it in the *x* direction. The ponderomotive force (pressure) acting on a unit surface area of the sphere is similar to a hydrostatic force and can be found from the field strength on the sphere, $E_0$. The same pressure acts on the area $s$, giving $T = -\pi(a^2 - x^2)\frac{\varepsilon_0}{2}E_0^2$.

Inside the sphere, the total nonelectromagnetic (mechanical) momentum is

$$p_i = \frac{v_x}{c^2}\int_{-a}^{a}T dx = -\frac{q^2}{32\pi\varepsilon_0 a^4}\frac{v_x}{c^2}\int_{-a}^{a}(a^2-x^2)dx = -\frac{1}{3}\frac{q^2}{8\pi\varepsilon_0 a}\frac{v_x}{c^2}. \qquad (9)$$

This additional momentum is negative and is exactly equal to a quarter of momentum (1). This yields the momentum of the charge sphere in the form

$$p = p_f + p_i = \frac{1}{c^2}\frac{q^2 v_x}{8\pi\varepsilon_0 a}$$

which is already in exact correspondence with Eqn (2). Remarkably, this result is independent of the distribution of stresses $t_{ij}$ within the sphere, or of what its interior is (inhomogeneous, hollow, anisotropic, liquid, and so on). We note that rather than producing the expected addition to the electromagnetic mass, the solution to the problem has led to an `extra' momentum, necessary for obtaining a closed solution.

8. In the special case of an isotropic sphere, the solution is extremely straightforward to obtain. From Eqn (8), noting that the volume of the sphere is $V = \frac{4}{3}a^3$ and the pressure on its surface is $p = -\frac{q^2}{32\pi^2\varepsilon_0 a^4}$, we calculate mechanical momentum (9) as the momentum density times the volume, yielding

$$p_i = \frac{v_x}{c^2}pV.$$

9. The general solution that gives the same result can be obtained for this paradoxical problem directly from the equivalence of momentum and energy density (4) using the expression for the momentum [8, 10]

$$\delta P = \frac{1}{c^2}\frac{d\mathcal{E}}{dt}l, \qquad (10)$$

where $l$ is the separation between the elementary sources and receivers of energy, and E is the energy. We note that the energy is generated and received by the sphere (see Section 6 and Fig. 1), and that the surface power density is

$$\mathbf{E} \cdot \mathbf{j} = \frac{\varepsilon_0 v_x}{2} E_0^2 \, a \sin \theta,$$

where $E_0$ is the field strength on the sphere and $\theta$ the angle between the $x$ axis and the radius of the vector starting at the center of the sphere. Summing the energy flows (10) over elementary surface areas and passing to an integral yields the same value of the moment as in Eqn (9):

$$p_i = -\frac{1}{c^2} \frac{q^2 v_x}{8\pi \varepsilon_0 a^2} \int_0^{\pi/2} \cos^2 \theta \sin \theta \, d\theta = -\frac{1}{3} \frac{1}{c^2} \frac{q^2 v_x}{8\pi \varepsilon_0 a^2}.$$

We note that the density of the medium is absent from expressions for the momentum of the medium, which is due to the stresses caused by ponderomotive forces. Because this medium plays the same role as the Poincare tensions, it can be argued that the classical model of the electron, as well as Poincare's hypothesis, cannot resolve the 4/3 paradox in principle.

Problems of this type show that general principles apply in special cases. Some instructive problems and some not necessarily obvious conclusions from considering the relativistic (hidden) mechanical momentum have already been discussed in the literature (see, e.g., Refs [11±14]). It should be noted that the motion of free charges can generally be described by considering the electromagnetic field momentum only (see the famous paper by Page and Adams [15]). But if the motion of charges is not free, we may face the necessity of taking the hidden momentum into account [16, 17].[3]

We acknowledge the support and helpful advice of MB Belonenko, GD Bogomolov, and Yu N Eroshenko.

---

[3] The author of Ref. [17] incorrectly called the paradox `the Onoochin paradox,' but it was actually proposed by G P Ivanov, as was noted in the Russian translation of Ref. [17].

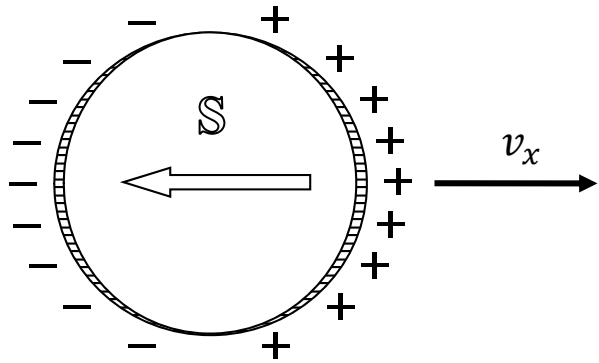

Figure 1. Energy flow $\mathbb{S}$ inside a charged sphere as it moves. The rigid sphere undergoing a displacement $dx$ `absorbs' the field $E_0$ on one of its sides and `creates' it on the other (hatched regions). The plus (minus) sign marks the region where the sphere takes in energy, $\mathbf{E} \cdot \mathbf{j} > 0$ (gives off energy, $\mathbf{E} \cdot \mathbf{j} < 0$).